\documentclass{article}
\usepackage[utf8]{inputenc}
\usepackage{float} 
\usepackage{graphicx} 
\usepackage{hyperref}
\begin{document}
\title{Considerations for Interdisciplinary Instruction and Design Influenced by Adaptive Learning. A Case Study Involving Biology, Computer Science, Mathematics, and Statistics.}
\date{\today}
\author{Karen Aguar, Charles C. Sanchez, Diego Boada Beltran, \\
Saeid Safaei, Mehdi Assefi, Jonathan Arnold, Pedro Portes, \\
Hamid R. Arabnia, Juan B. Gutierrez\footnote{Corresponding author: \href{mailto:jgutierr@uga.edu}{jgutierr@uga.edu}} \\ University of Georgia}
\maketitle

\begin{abstract}
ALICE (Adaptive Learning for Interdisciplinary Collaborative Environments) is an open-source web based adaptive learning system designed for interdisciplinary instruction. ALICE has the potential to transform education by empowering transdisciplinary knowledge acquisition. This is particularly important in fields that accept newcomers with diverse scholastic backgrounds, e.g. Systems Biology. With traditional interdisciplinary instruction, the instructor must cover pre-requisite information from multiple disciplines to ensure all students begin at a common baseline - slowing the learning process. With ALICE, students follow a personalized syllabus based on their previous knowledge and work towards individual goals. Implementing an adaptive learning system in an interdisciplinary course requires careful considerations of the instructional design. Structuring material, formulating assessments, and other instructional design aspects must be carefully considered. These considerations are detailed through the exploration of a case study implementing ALICE in a graduate level Systems Biology course. 
\end{abstract}

\section{Introduction}

	The dominating paradigm in interdisciplinary STEM education is inherently inefficient particularly for students from various disciplinary background attempting graduate studies. It consists of essentially teaching the same knowledge base to each student within the classroom; however: (i) students in these settings usually come from different disciplines, thus having different (often non-overlapping) backgrounds, and (ii) curricula in interdisciplinary fields are comprised by subject matter drawn from different (often traditionally disconnected) areas. Case in point, systems biology; in this area, students need to master a biological problem, know the theory of dynamical systems (continuous and discrete), probability, statistics, and be able to program, just to mention a few subjects. Students who take this interdisciplinary class at the senior undergraduate and junior graduate levels generally major in genetics, biochemistry, horticulture, mathematics, computer science, statistics, physics, engineering, etc. As a result of these multi-disciplinary skill requirements and the inherent diversity of student backgrounds in an interdisciplinary class, some students in the classroom have expertise in some areas and deficits in others, and these strengths and weaknesses are unique for each student. While it is possible to require all students in these settings to master simultaneously a collection of disciplinary content areas, the delivery of instruction in which all learn the same at equal pace poses uneven and unreasonable demands on students. Ideally, each student should strengthen her/his specifics weaknesses, and broaden the scope of their strengths within the same time frame allotted for the  class. 

	We developed an open-source Web-based cyber-learning tool that allows any team of instructors spanning several scientific disciplines to curate a constellation of interdisciplinary learning resources for the purpose of creating individualized or small group learning progressions for developing prerequisite competencies and responsive education to all students. The personalization of the learning plan or syllabus for each student depends on previous knowledge and individual learning goals. This customization is achieved through an information system called ALICE (Adaptive Learning for Interdisciplinary Collaborative Environments), which connects a series of atomic units of knowledge (termed lexias) though a dynamic path and presents it to the student for the purpose of acquiring a set of competencies. The metaphor of the tree is replaced in ALICE by a dense rhizome-like network that does not privilege a particular path, but instead offers a milieu for traversal. In practice, it is the student during the learning process who makes an abstract knowledge network come to a unique realization. ALICE was initially designed for graduate and senior undergraduate learners in the subject matter of Systems Biology. Based on task analyses and dynamic assessments of individual learners, each learning progression was designed to take the learner from individual baselines to desired levels of competence. 
    
    ALICE personalizes education by: (1) creating a knowledge map of course material that is unique for each student (i.e, a personalized syllabus); (2) suggesting individualized paths across the knowledge map based on student competencies/accomplishments; (3) providing accessible Web-based interfaces for students and instructors for storing and presenting class materials, for assessment, and for recording student paths; (4) establishing social networks for collaborative learning of course material through shared problem-solving tasks and for passing the research learning experience from current students to future ones. 
 
\section{Background}

	Adaptive learning is the notion of using computers as interactive teaching devices to adapt to the user's individual needs. It combines the fields of Computer Science, Education, Psychology, etc. The computer adapts the way it presents material or decides what the next question will be based on its interactions with the students - via observing the student and or analyzing their responses. Adaptive Learning is a broad term that encompasses many Adaptive Learning System (ALS) varieties such as Adaptive Educational Hypermedia (AEH), Intelligent Tutoring Systems (ITS), Adaptive eLearning, and others. It is primarily used in educational settings such as classrooms and business training \cite{aguar2016inclusive}.

	Research on Adaptive Educational Hypermedia (AEH) has gained significant interest in the last two decades. Adaptive hypermedia systems are built on a model of the goals, preferences, and knowledge of individual users to adapt to their specific needs \cite{brusilovsky1996methods,brusilovsky2007user}. Such systems modify learning experiences on the basis of the system's ability to identify the learner's needs (i.e. adaptivity) and the possibility for learners to make decisions on their own (i.e. adaptability). The majority of systems to date have addressed adaptivity based on learning styles or cognitive models \cite{akbulut2012adaptive}. Frameworks such as the Felder-Silverman learning style dimensions \cite{felder1988learning}, Keefe's classification of learning styles \cite{keefe1987learning}, and cognitive styles models \cite{pask1976styles,gardner2000intelligence} have often guided the design and implementation efforts of AEH systems. Evidence suggests that AEH systems are effective at tailoring instruction for heterogeneous groups of students both in higher education \cite{van2012influence,fullilove1990mathematics} and in K-12 settings \cite{chen2014adaptive,walkington2013using}.

	Adaptive learning systems alter a student’s learning experience based on the system’s perception of the learner’s needs. The study of learning styles and cognitive models have guided the development of the majority of systems to date \cite{akbulut2012adaptive}. Recent studies of adaptive systems report improvements in perceived learning, motivation, and overall satisfaction and  learning experience \cite{buch2001accommodating,chen2014adaptive,lo2012designing,mampadi2011design,papanikolaou2003personalizing,peredo2011intelligent,sangineto2008adaptive,taylor2007adapting,triantafillou2003design,triantafillou2004value}. A comprehensive review of the literature can be found in Akbulut \& Cardak (2012) \cite{akbulut2012adaptive} and Somy\"urek (2015) \cite{somyurek2015new}. A mathematical intelligent tutoring system designed by Arroyo et al (2014) \cite{arroyo2014multimedia} found positive impacts on students’ cognitive, affective and metacognitive abilities. Both short-term and long-term learning benefits were demonstrated by pre and post-test measures and state-wide standardized test scores along with significant increases in motivation levels, engagement, and other affective outcomes. Van Seters et al. (2012) \cite{van2012influence} used adaptive e-learning materials in a graduate level molecular biology course in Europe where students had diverse backgrounds and widely varied prior knowledge. They argue that these varying factors require intensive tutoring and scaffolding. We ought to mention that this effort was mono-disciplinary. 

	However, it is still somewhat unclear what makes adaptive systems effective and how they affect students' learning aptitudes, motivation and performance \cite{van2012influence,chen2014adaptive}. A limitation is that adaptive systems typically rely on a single-source of personalization of information such as learning style, cognitive style, prior learning achievement, or motivation \cite{van2012influence,chen2014adaptive} that lack a strong research base.

\section{ALICE} 
\begin{figure}[H]
    \centering
    \includegraphics[width=\linewidth]{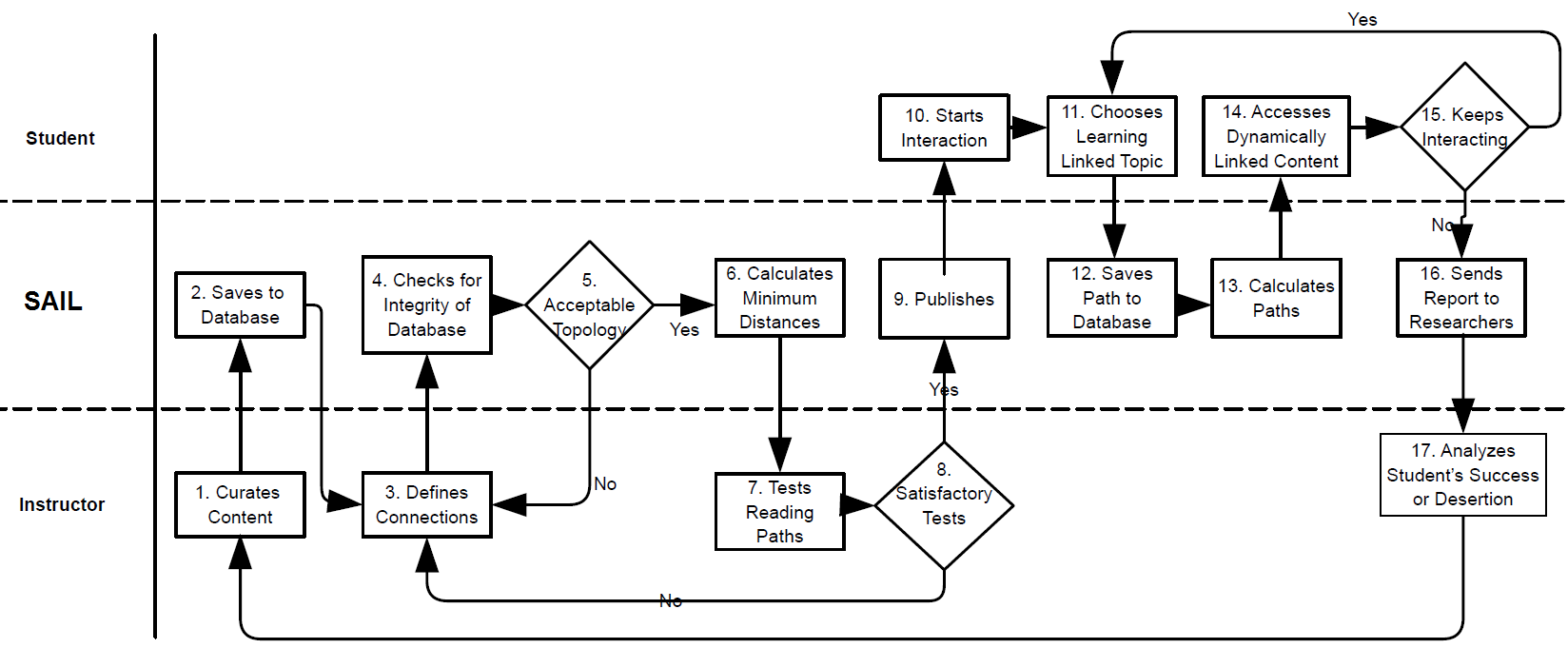}
    \caption{The flow of information in ALICE has three interconnected domains: students, the information system, and the instructor. ALICE plays a fundamental and autonomous role in guiding the students through the material.}
    \label{fig:ALICEflow}
\end{figure}

The architecture of ALICE is based on the Literatronica system \cite{gutierrez2006literatronic,gutierrez2008literatronica}. Figure \ref{fig:ALICEflow}  shows the workflow that permits adaptive behavior. This optimization process is achieved through the real-time solution of a multi-terminal network flow and maximal network flow on a dynamic graph.   In ALICE each competency has a determined finish point. Each time a learner interacts with the system, ALICE reconfigures links to have different destinations, leading every time to a personalized and potentially unique learning experience. ALICE offers an adaptive behavior that ranges from maximal flow (i.e. completion of the track with the maximum number of lexias) to minimal path (i.e. completion of the track with minimum number of lexias). 

\begin{figure}[H]
    \centering
    \includegraphics[width=\linewidth,keepaspectratio]{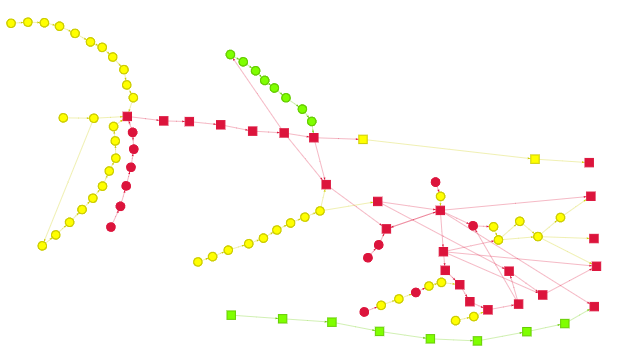}
    \caption{The ALICE knowledge map for the Systems Biology pilot. Yellow icons represent mathematics and statistics lessons, green represent computer science, and red represent biology. }
    \label{fig:ALICE_knowledge_map}
\end{figure}

\begin{figure}[H]
    \centering
    \includegraphics[width=\linewidth,keepaspectratio]{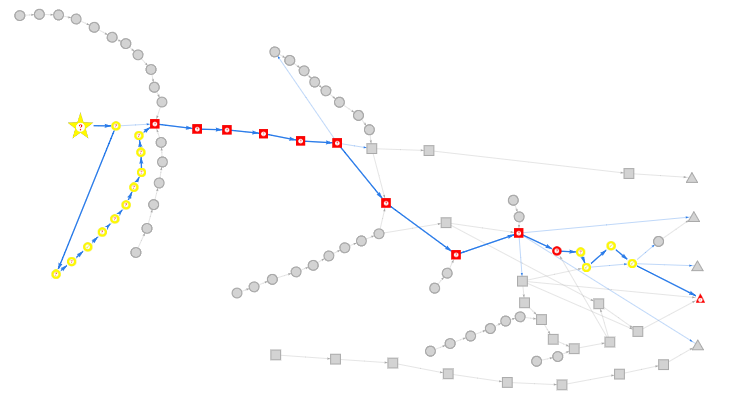}
    \caption{The above image shows a student's unique path through the knowledge map to reach their selected capstone experience. Items in gray are not included in the student's IDP.}
    \label{fig:ALICE_unique_path}
\end{figure}

    ALICE presents a knowledge map to the students - as illustrated in Figure \ref{fig:ALICE_knowledge_map}. Colors identify the main area of each lesson - yellow represents mathematics and statistics, red represents biology, and green represents computer science. Square shapes represent lessons related to the central theme of the course, while circles represent pre-requisites and triangles represent the five capstone experiences: (i) single cell clocks, (ii) circadian rhythms, (iii) host-pathogen interactions, (iv) development of vascular and lymphatic networks,and (v) thymus development. Once students select a capstone experience, their first lesson is identified with a star. Figure \ref{fig:ALICE_unique_path} shows an example of what a student's IDP might look like to reach their selected capstone experience.
    
%
%
    

    

	With ALICE, the traditional lecture-style classroom is not a viable option as students will gain competencies in different areas at different paces. Below we outline the changes in course design necessary to incorporate an adaptive learning environment. 
    
\section{The Collaboratorium}
    
	ALICE represents a substantial change in paradigm with respect to traditional instruction. It requires a teaching approach or model that accommodates interdisciplinarity and heterogeneity of skills, while ensuring that students acquire desired competencies. What is normally missing from traditional curricula is the chance to engage with science as it occurs in practice, where members of a scientific project team have particular strengths and collaborate to achieve a common goal. We used ALICE following the model of a collaboratorium, in which students are part of an interdisciplinary collaborative team asked to carry out a research project, i.e. capstone, successfully in what has been termed problem–based learning. We define a collaboratorium as a research-centered course spanning multiple disciplines.This collaboratorium facilitates forming heterogeneous small groups of students during and after classes, on site and online to assist students through what has been coined dialogical learning \cite{corno2015handbook}. 

\begin{figure}[H]
    \centering
    \includegraphics[width=3in,keepaspectratio]{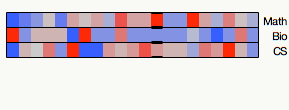}
    \caption{Student Competency in Math(and Statistics)/CS/Bio in Spring 2017 Pilot Pre-assessment. Each column is the competency profile of one student. Red indicates a higher competency in each skill.}
    \label{fig:pre_post_competency}
\end{figure}

     Students come into the collaboratorium with very diverse backgrounds. For example, in Systems Biology some students may be mathematicians, some statisticians, some computer scientists, and some biologists. Figure \ref{fig:pre_post_competency} shows a heat map of students' strengths and weaknesses in each of these subjects at the beginning of the ALICE-SB (Systems Biology) pilot study (Spring 2017). Each student may have expertise in one area but may need to come up to speed in the other areas. One of the major challenges of a collaboratorium is an intelligent structuring of this interdisciplinary environment. While a collaboratorium could exist without ALICE, it is the interdisciplinary medium in which ALICE operates. With ALICE, each student does not have to master all concepts pursued in the collaboratorium, just enough to ensure that the collaboratorium succeeds. 
     
	ALICE enables the ability to expand adaptively according to what the students need to learn, saving instructors the expense of energy and time of teaching students what they already know. The hypothesis is that ALICE will allow students to achieve higher competency gains in the areas they are less familiar with since they are not re-learning material they know. This process accelerates the process with which the students gain expertise in other areas. 

	Instructors are drawn from multiple disciplines and function as a collective student resource. Other interdisciplinary teaching approaches have been used in higher education yielding positive outcomes  \cite{bucci2014children, fox2014mix, fuselier2015teaching, russell2015}. For example, the Mix concept and design at the University of British Columbia \cite{fox2014mix} consisted of regularly scheduled workshops for faculty and staff to cultivate an interdisciplinary community and explore potential partnerships. As a result, faculty developed joint group student projects, interdisciplinary lectures, field trips, "data mash-ups", and electronic discussions across courses, all of which engaged students with communities both within and outside the University. However, little is known about the factors that facilitate academic boundary crossing, overcome barriers to successful curriculum transformation, and enable the usefulness of technology-enhanced environments. We are not precluding that ALICE is absolutely necessary with such a diverse community, but hypothesize ALICE to be a technological contribution to the development of inter- and multi-disciplinary ability among students in the sciences. 
    
	We anticipate that ALICE can help develop a robust emphasis on cross-cutting concepts and interdisciplinary collaborations in science education by providing a means for instructors and curriculum developers to plan lessons and courses tying together core ideas in ways that activate student learning and at the same time promote cognitive strategies in self-regulated learning and metacognition \cite{pashler2007organizing}. For more homogeneous subjects, we still think ALICE could serve as an improvement for ensuring adequate pre-requisite knowledge. This instructional model would take a different structure from that of heterogeneous groups, both are described below. 
    

    
\section{ALICE Influence on Instruction and Design}

	\subsection*{Course Plan}

	Preparing for a course with ALICE involves the same careful planning of any course.  Activities of constructing a syllabus, handouts, and lectures have their counterparts in an ALICE-assisted class.  The teachers or project leaders must prepare: (1) a knowledge map of all lexias to be covered (\textit{i.e}, a syllabus); (2) short write-ups for each lexia to be covered to be uploaded to ALICE (\textit{i.e}., a handout);(3) lectures to be loaded into ALICE.
	In order to prepare the knowledge map, for each lexia or topic the competencies to be gained must be identified for each lexia as well as the prerequisites for mastering the lexia.  These items are loaded into ALICE.  As an example, one lexia in the Systems Biology course is "a genetic network".  The competency gained can be framed as a question, "Can you simulate a genetic network?".  In order to acquire this competency, certain prerequisites must be met by knowing what is a: (1) gene; (2) gene cluster; (3) derivative; (4) differential equation; (5) limit; (6) kinetic model; (7) transcription; (8) translation.  All of the prerequisites can be handled in ALICE by loading them as Wiki pages for access by the students. The instructor writes the handouts and the lectures as usual.
    Once the knowledge map is created and populated, the topics must be linked up.  In a syllabus they are linearly arranged. In a knowledge map the arrangement of topics is a little more complicated, allowing greater flexibility about how a student progresses.  Each lexia is linked to at least one other lexia.  The link is assigned a distance value ranging from 1 to 5.  We suggest a link value of 1 for two lexias or topics that are closely related, and we suggest a link value of 5 for two topics that are only distantly related.  These distances between lexias are the instructors' assessment of how the lexias are related in a network of topics. Prerequisites are handled in the same way based on the examination of the content of the wiki pages.  The student's path through the lexias to the capstone experience is the individualized syllabus or individualized development path (IDP).  Depending on a student's background, different students will have different paths through the course.  In Figure \ref{fig:ALICE_knowledge_map} is the knowledge map and distance values assigned to the course lexias.  There are ~20 lexias in this knowledge map.  As the course is finalized, we expect about 30 lexias in the course.

    As the students progress, they will upload their homeworks to ALICE.  The instructors comment on them in ALICE to determine their competency.  In that graduate students are expected to have a B in a graduate course to obtain credit toward their degree for many programs at the University of Georgia, we decided that a grade of 80 or above would then translate into competency achieved in a particular lexia.  What is a little different about ALICE-based classroom is there is no set order to assignments, as students complete different lexias on their IDP to the capstone experience. Decisions regarding homework deadlines, assessments, and course progression are explored in section \ref{homework} below.

	\subsection*{Adapting the Classroom}
    \label{adapting}
	With adaptive technology providing students a unique path and learning experience throughout a course, classroom lecture meetings must also adapt. The traditional design of the instructor delivering content, i.e. lectures, during class time is extinct as students gain their personalized lectures via ALICE. Many ALS and AEH systems have been geared towards online classes for this reason. Although the traditional classroom setting is irrelevant, we argue that to eliminate borders in an interdisciplinary field and to promote deeper learning, in-class collaboration is necessary. With ALICE, the classroom becomes a discussion-based setting to encourage collaborative thinking and problem solving skills between disciplines. 

	The challenge is that students are each on their IDP, therefore classroom discussions must be created carefully. Specific exercises geared towards a particular domain are not useful in this setting, but broad, multi-disciplinary problems (or case-studies) that can help students explore solutions as a group from many different angles are the key. For example, students can be encouraged to give ongoing reports on their capstone experiences to the rest of the class beginning with a presentation early in which they describe how they will get to the capstone experience. Students already equipped with some competency, such as programming or the construction of genetic networks, can be encouraged to share their expertise with students within their project group as defined by the capstone experience. 


	With ALICE, the instructor's role changes from giving a lecture to answering questions, encouraging progression, providing feedback, and facilitating discussions. It is important that no centralized lecture be given at the start of class, but instructors should monitor where students are in lexia progression prior to class time to facilitate collaboratorium conversations. In every class instructors can visit each student to ascertain where they are having problems. It may be in understanding transcription or how to build a bifurcation plot in the analysis of dynamical systems. Capstone experiences, or group projects, are the main driver of progression for the ALICE students - so it is important that these are chosen to be interesting, research-backed problems to motivate students. The engine for class progress is student enthusiasm for a particular capstone experience. 

	In the case that the class is more homogeneous, ALICE is still suggested as an improvement to traditional instruction as it can help fill pre-requisite knowledge gaps. This classroom structure would differ from the typical heterogeneous group described above. Instructors in homogeneous groups may want to provide a short, centralized lecture at the start of class and use ALICE to further readings, fill knowledge gaps, and monitor student progression through homeworks. In this setting, students would progress in a more linear fashion with same starting and ending goal but with some variation in between based on pre-requisite knowledge.
    
    	\subsubsection*{Homework and Progression Considerations}
       	\label{homework}
        
	ALICE is designed so that students will complete short lessons to gain new knowledge or lexias throughout the course. Each lesson can consist of powerpoints, PDFs, videos, live-scribe lectures, or any other kind of multimedia determined useful by the instructor for learning. After completion of each lexia, the student is required to complete a terminal homework assignment (determined by the instructor) for the topic to prove competency in that area. Students upload their assignment to ALICE, and are then allowed to progress to the next lexia in their IDP. The decision was made to allow students to progress before an assignment was graded so they would not have to wait on instructor feedback to move forward with knowledge gains. 
    
    Since ALICE creates unique learning paths for each student, students will be completing different assignments as they progress on their IDP towards the capstone experience and some may complete more lessons than others throughout the duration of the course. With such flexibility in assignments, the following questions must be considered when designing a course:
    \begin{itemize}
    	\item What is the minimum number of lessons a student should complete?
        \item How is student progression through lessons encouraged?
        \item How is grading lessons handled?
	\end{itemize}
      
    To decide on the minimum number of lessons each student should complete throughout an adaptive course, a thorough examination of the design of content in the course should be conducted. For our Systems Biology course, the majority of students come from one of three predominant backgrounds: mathematics and statistics, computer science, or biology. The compilation of lessons created by the instructor consisted of approximately one-third mathematics based lessons, one-third computer science and one-third biology. It is expected that students will have a strong background in at least one of the three subjects, eliminating one-third of the possible lessons from their unique path. It is also possible that students come in with a strong background in two of the three disciplines, perhaps with strong mathematics/statistics and computer science skills but little biology knowledge. Such a student’s IDP would likely consist of only the third subject’s content, or about one-third of the total number of lexias in the entire course – this determined our baseline for the minimum number of assignments or lessons that a student should complete. If a student is competent in more than two-thirds of the content for the course, they likely would not be taking the class. Thus, we require all students to complete at least one-third of the total assignments for the “homework” portion of their grade.

	Once a minimum number of lessons for your course is determined, instructors will need to consider how to grade these assignments, how to address those students completing more than the minimum number of required competencies, and how to encourage or enforce lexia progression throughout the course. Since the System Biology course is not required for a major, students enrolled tend to have high intrinsic motivation for their individual learning outcomes in the course. Due to this high intrinsic motivation, students tend to be more driven to gain competencies autonomously. We decided not to enforce progression throughout lessons on a predetermined schedule or compensate students more/less for completing more than the minimum required assignments in our pilot study and will report our findings once the experiment has concluded. If an adaptive course requires more substantial extrinsic motivating factors to encourage student progression and completion of lexias, policies should be adjusted accordingly.

    	\subsubsection*{Capstones}
	We suggest designing capstone experiences, i.e. group projects that encourage students to work towards a research project in heterogeneous groups. Throughout the semester, they should submit reports on their progress along the way. This interdisciplinary collaboration allows students to work together in a real-world research environment where students have differing strengths and weaknesses to accomplish a united task. We suggest allowing students to define the goals of their own research project (while adhering to necessary requirements for the course) to boost intrinsic motivation and self-motivated in their projects.  Part of the weekly reports early on in a class can be presenting these learning goals to the class.
    
   Students should be encouraged in the first weeks of class to form heterogeneous study groups on various projects determined  by areas they are most interested in. Forming groups with diverse learning or disciplinary background information is processed by the instructor early on when student interests in selected projects. As detailed elsewhere, as each students' profile is the most important variable in this approach, knowing about students' prior learning experiences for teaching can promote intrinsic motivation throughout the course. Additionally, the capstone experiences that require writing performance allow instructors to assess students' cognitive understanding of the problems in class along the educational objecting found in Bloom's Taxonomy \cite{bloom1956taxonomy}.For example, using various forms of assessments formatively can help guide collaboratorium activities that help students to go from comprehension to analyzing, synthesizing and later on evaluating.

    	\subsubsection*{Assessment}
        
        In order to achieve the above goals, assessments at the beginning of the course are required to determine a student's current strengths and weaknesses. This is needed so that students may ultimately navigate their own unique paths throughout the course. The first goal is to determine what pre-requisite knowledge each student possesses and what knowledge they need to gain. This can be determined through individual interviews, which may be extended in smaller class groups or a pre-assessment exam, though this would need to be carefully crafted. In either case, the goal is to ask students questions about both correct and incorrect responses that could accurately determine where their path should start in the ALICE system and what additional content needs to be included in planning after and in-class discussions towards integration and synthesis in shaping their own capstone experience and products. When a written or digital pre-assessment exam is used, instructors should be careful to craft each question in a meaningful way to assess accurately knowledge levels, which is challenging. Multiple choice for diagnostic purposes are difficult items to craft, as students may be able to guess correctly otherwise and therefore exempt lexia competencies from being activated in their learning path. Students should be strongly encouraged not to guess and reassured that however they are assessed, it is for their benefit and will not count against them in any way. 

	At the same time, discussion about why wrong choices are selected in these multiple choice or True/False questions can be valuable in planning the use of lexias and  mastery of competencies in achieving course educational goals. The latter are defined by what students should be able to do after each lexia as the unique paths are completed  in the learning progressions  organized for the class. A good way to craft such questions is to become familiar with tips  from instructional psychology and design resources in selecting test bank questions in the disciplinary areas or books integrated in the bioinformatics course (3 areas) and then see which can be improved for the current class.  For example, when analyzing or writing a test item, present an authentic problem that requires one or more of the rules for analysis in problem solving or one that requires discrimination between two competing concepts, such as saddle point bifurcation and trans-critical bifurcation.
	Assessment involves using an item stem that presents a problem that requires application of course lexia knowledge or analysis of a problem, by evaluating carefully worded alternatives focused on testing higher-order thinking and students’ ability. To evaluate or develop multiple choice items to test higher order thinking, at least one of the choices besides the correct answer must be plausible but designed to detect a critical gap. The other choices or distractors can be less difficult.

    As students are learning different background knowledge to solve the same problems, choosing to have exams throughout requires special considerations. Exams should not test only lexias, or surface-level knowledge, but should test application of content and the development of deeper problem-solving skills based on Bloom's Taxonomy \cite{bloom1956taxonomy}. Open-ended questions that allow students to demonstrate mastery from a certain angle can be effective here.  For example, why do you use ensemble methods in model identification in systems biology or how would you use a metabolomics approach to identify a mating pheromone in a worm?

    Choosing to have exams at specific times in the semester will mean requiring students to have mastered a certain amount of content (lexias) by an exam deadline. This can be a useful option to ensure that students gain sufficient knowledge at these points during the semester. To ensure students preparedness, giving a practice exam may be a good option to allow students to anticipate the kinds of questions asked and the level of knowledge expected for the exams - this can also give the instructor a chance to test the way they have crafted their questions.  If overall competency is to be assessed through a test, the questions of the test must reflect the diversity of paths taken by students through the knowledge map.  A test can be constructed on which students are assessed on only a subset of questions.  For example, ten questions might be framed on the test, and each student is only required to answer three questions from lessons traversed in the knowledge map.

\section*{Future}

	To begin exploring viability and assess ALICE’s impact, a pilot study in a Systems Biology course is being conducted in the Spring 2017 semester. Two sections of Systems Biology is being offered and contrasted in various ways. One section  serves as the “control” group and is  taught in a traditional lecture-style classroom setting exactly as it has been taught in the past using best current practices. The other section is the “experimental” group and uses ALICE-SB for mediating learning and teaching. Systems Biology is an ideal interdisciplinary course combining skills from biology, mathematics/statistics, and computer science. Students taking this course come from a variety of disciplines and represent a diverse population of educational levels ranging from advanced undergraduates to post-doctoral fellows. The interdisciplinary nature of this course makes it an excellent candidate to test ALICE. 
    
	The self-assessment process we intend to use is that codified by McMillan and Hearn (2008), a cycle of self-judgment, setting learning goals, and self-monitoring as students who practice self-assessment tend to persist through learning moments and tasks that might otherwise be overwhelming (Rolheiser, 1996). This is critical as technology-assisted learning tends to suffer from ‘novelty effects’ (Cheung \& Slavin, 2011; Li \& Ma, 2010), early spikes in technology use and academic improvement followed by regression to prior levels of academic proficiency. Fortunately, ALICE is not designed as a standalone tool and is envisioned as part of a hybrid approach to education. A recent meta-analysis (Means, Toyama, Murphy, Bakia, \& Jones, 2010) finds that a mixed approach generally yields greater results than independent or classroom-only learning.

Specifically, the research questions we hope to answer in the future are:
\begin{enumerate}
\item What sorts of changes, if any, does using ALICE create in instructional planning and design?
\item To what extent does ALICE’s unique adaptive features contribute to the removal of barriers to learning cross-cutting concepts in inter- and trans-disciplinary environments? 
\item How does ALICE promote the development of novel problem-solving skills in its users? 
\item To what extent does ALICE remove barriers to collaboration and communication among students and between students and instructors? 
\item Does ALICE contribute to motivating interest in learning for varying populations of students? 
\end{enumerate}

At this time, we have identified many necessary changes in instructional design required to incorporate ALICE and have documented these requirements in this paper.

\section*{Conclusion}
	We have created ALICE (Adaptive Learning for Interdisciplinary Collaborative Environments), an open-source web-based cyber-learning tool that allows personalization of the learning plan or syllabus for each student depending on previous knowledge and individual learning goals. ALICE is an ideal system for interdisciplinary settings where students come from a variety of backgrounds, as it eliminates redundant information and allows each student to strengthen her/his specific weaknesses, and broaden the scope of their strengths within the same time frame allotted for the class. Within ALICE, instructors spanning several scientific disciplines can curate a constellation of interdisciplinary learning resources for the purpose of creating individualized or small group learning progressions for developing prerequisite competencies and responsive education to all students. ALICE significantly impacts the role of the instructor and design of the course. 

	As we pilot ALICE in a graduate level Systems Biology course, ALICE-SB, we report consideration and modifications in instructional design needed to implement ALICE. The architecture and guidelines to incorporate ALICE in your own course are outlined. The traditional lecture style is no longer possible as students acquire knowledge personalized to their individual background and goals. With ALICE, the classroom evolves into a discussion based setting to encourage collaborative thinking and problem solving between disciplines. We suggest the use of a collaboratorium – where heterogeneous groups of students work together to solve interdisciplinary research problems. The instructor’s role changes from lecturer to facilitator of learning. Normal course elements such as centralized assessment techniques, due dates, course progression, and grading schemes must all adapt with personalized instruction. The influence of ALICE on these instructional design aspects are detailed and discussed.

\section*{Materials and Methods}
The pilot course was conducted at the University of Georgia (UGA) during spring, 2017. The quantitative results of the pilot will be published in a subsequent publication. The  recruitment rate was all students but 1 in two classes of size 13 and 9 (BINF 8950 advanced graduate level, and MATH4780/MATH6780 senior undergraduate or beginning graduate levels), one with ALICE and one without ALICE respectively.  Data on participants was collected under Institutional Review Board approval and deposited either in the ALICE database or the eLearning Commons database at UGA.  Both databases are password protected and behind the University firewall.  Participant data included: (1) test results; (2) homeworks on each lexia; (3) personalized syllabi for each student.  Test results were followed up with one on one meetings with participants to validate the test results.
\section*{Acknowledgment}
This material is based upon work supported by the National Science Foundation under Grant No. 1645325. 


\end{document}